\documentclass{article}

\usepackage{PRIMEarxiv}
\usepackage{comment}
\usepackage[utf8]{inputenc} 
\usepackage[T1]{fontenc}    
\usepackage{hyperref}       
\usepackage{url}            
\usepackage{booktabs}       
\usepackage{amsfonts}       
\usepackage{amsmath}        
\usepackage{nicefrac}       
\usepackage{microtype}      
\usepackage{fancyhdr}       
\usepackage{graphicx}       
\usepackage{multirow}       
\usepackage{rotating}       
\usepackage{makecell}       
\usepackage{xcolor}         
\usepackage{svg}
\usepackage{graphicx}
\usepackage{tablefootnote}
\usepackage[most]{tcolorbox}
\usepackage{enumitem} 

\graphicspath{{media/}}     

\title{Whisper Leak: a side-channel attack on Large Language Models}

\author{
  Geoff McDonald \\ 
  Microsoft \\ 
  Vancouver, Canada \\ 
  \texttt{geofm@microsoft.com} \\ 
   \And
  Jonathan Bar Or (JBO) \\ 
  Microsoft \\ 
  Redmond, USA \\ 
  \texttt{jobaror@microsoft.com} \\ 
}

\begin{document}
\maketitle

\begin{abstract}
Large Language Models (LLMs) are increasingly deployed in sensitive domains including healthcare, legal services, and confidential communications, where privacy is paramount. This paper introduces Whisper Leak, a side-channel attack that infers user prompt topics from encrypted LLM traffic by analyzing packet size and timing patterns in streaming responses. Despite TLS encryption protecting content, these metadata patterns leak sufficient information to enable topic classification. We demonstrate the attack across 28 popular LLMs from major providers, achieving near-perfect classification (often >98\% AUPRC) and high precision even at extreme class imbalance (10,000:1 noise-to-target ratio). For many models, we achieve 100\% precision in identifying sensitive topics like "money laundering" while recovering 5-20\% of target conversations. This industry-wide vulnerability poses significant risks for users under network surveillance by ISPs, governments, or local adversaries. We evaluate three mitigation strategies - random padding, token batching, and packet injection - finding that while each reduces attack effectiveness, none provides complete protection. Through responsible disclosure, we have collaborated with providers to implement initial countermeasures. Our findings underscore the need for LLM providers to address metadata leakage as AI systems handle increasingly sensitive information.
\end{abstract}


\section{Introduction}

Large Language Models have rapidly become essential tools for information retrieval, content generation, and decision support across diverse domains. As LLMs handle increasingly sensitive data - from medical consultations to legal advice to private conversations - ensuring confidentiality of user interactions is critical. While TLS encryption protects communication content from eavesdroppers \cite{satapathy2016comprehensive}, metadata such as packet sizes and timings remain observable, creating potential side-channel vulnerabilities \cite{hettwer2020applications, zhang2024timing}.

Recent work has identified several LLM-specific side channels: token length leakage from packet sizes enabling response reconstruction \cite{weiss2024your}, timing variations from efficient inference techniques revealing prompt properties \cite{carlini2024remote}, output token counts correlating with sensitive attributes \cite{zhang2024time}, and cache-sharing timing differences exposing input content \cite{zheng2024inputsnatch}. These attacks exploit fundamental characteristics of LLM deployment: autoregressive generation, streaming APIs for responsiveness, and encryption properties that preserve size relationships.

\textbf{This work.} We present Whisper Leak, a side-channel attack that infers user prompt topics by analyzing encrypted network traffic patterns during streaming LLM responses. Unlike prior work targeting response reconstruction or specific optimization artifacts, our attack classifies the high-level topic of conversations by learning patterns from sequences of packet sizes and inter-arrival times across diverse prompts. Our work introduces additional work that puts models at risk despite defenses such as token batching.

\textbf{Threat model.} We consider a passive network adversary - an ISP, government agency, or local network observer (e.g., coffee shop WiFi) - who can monitor encrypted traffic but cannot decrypt it. The attacker's goal is to identify when users discuss specific sensitive topics (e.g., political dissent, regulated activities, health conditions) among general background conversations. This scenario is particularly concerning for vulnerable populations in restrictive environments where mere discussion of certain topics carries risk.

\begin{figure}[htbp]
    \centering
    \includegraphics[width=\textwidth]{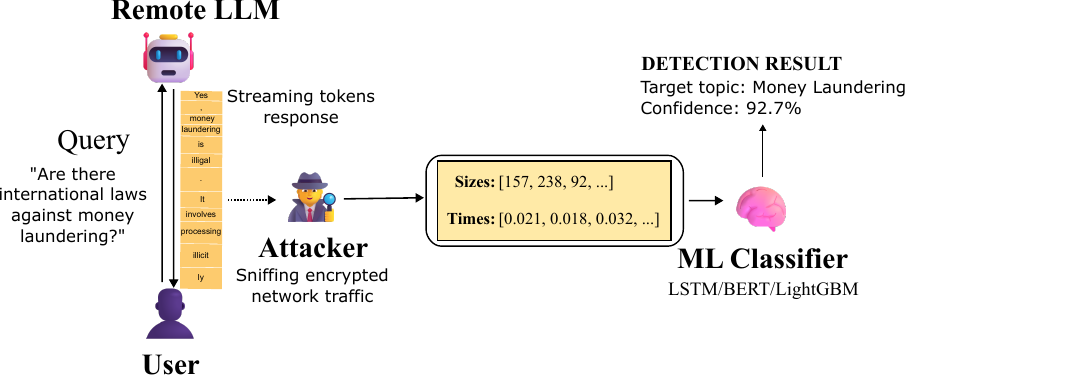}
    \caption{Whisper Leak attack pipeline: A passive network adversary observes encrypted TLS traffic between user and LLM service, extracts packet size and timing sequences, and uses trained classifiers to infer whether the conversation topic matches a sensitive target category.}
    \label{fig:system}
\end{figure}

\textbf{Methodology.} We evaluate 28 commercially available LLMs from major providers by collecting encrypted traffic for up to 21,716 queries per model: 100 variants of a question on a sensitive target topic ("legality of money laundering"), mixed with 11,716 diverse questions from the Quora Questions Pair\cite{quora_questions} dataset. We train binary classifiers (LightGBM, LSTM, BERT-based) on packet size and timing sequences to distinguish the target topic from background traffic. Figure \ref{fig:system} illustrates our attack pipeline.

\textbf{Key findings.} Our results reveal an industry-wide vulnerability:
\begin{itemize}
    \item \textbf{High attack effectiveness:} For most models, classifiers achieve >98\% AUPRC, with many reaching near-perfect performance using packet sizes alone.
    \item \textbf{Practical precision:} Under more realistic 10,000:1 noise to target question imbalance, 17 of 28 tested models enable 100\% precision at 5-20\% recall, allowing adversaries to identify target conversations with minimal false positives.
    \item \textbf{Scaling concerns:} Attack performance improves substantially with more training data (Figure \ref{fig:dataset_size_compact}), suggesting real-world adversaries could achieve higher effectiveness.
    \item \textbf{Partial mitigations:} Token batching, packet injection, and random obfuscation each reduce but do not fully eliminate attack risk, highlighting the difficulty of defending against metadata leakage while preserving streaming performance.
\end{itemize}

\textbf{Contributions.} We (1) identify and demonstrate a novel topic inference attack exploiting streaming LLM traffic patterns; (2) evaluate 28 models across major providers, revealing systemic vulnerability; (3) estimate attack effectiveness under realistic adversarial conditions; (4) assess three mitigation strategies, characterizing security-performance tradeoffs; and (5) engage in responsible disclosure with providers, contributing to initial countermeasures.

\section{Background}
\label{sec:background}

Understanding the Whisper Leak attack requires background knowledge on how LLMs communicate and how the underlying encryption protocols function.

\subsection{Large Language Models and token streaming}
\label{sec:llm_streaming}
LLMs generate responses by predicting subsequent tokens (words or sub-words) based on the input prompt and the tokens generated so far. This process is inherently sequential or autoregressive \cite{zhang2024time}. Instead of generating the entire response at once, which could lead to significant delays, LLMs typically compute and produce tokens one by one. Efficient inference techniques like speculative decoding \cite{carlini2024remote} and cache sharing \cite{zheng2024inputsnatch} are sometimes employed to speed up this process, but can introduce data-dependent timing variations.

To enhance user experience and provide immediate feedback, many LLM applications employ *streaming*. As tokens are generated, they are sent immediately or in-batch to the client over the network, allowing the response to appear incrementally. This streaming behavior \cite{weiss2024your}, combined with the model's internal processing (e.g., attention mechanisms, potential Mixture-of-Experts architectures, key-value caching \cite{zheng2024inputsnatch}), influences the timing and size of the data chunks transmitted over the network.

\subsection{TLS encryption}
\label{sec:tls}
Communications with LLM services over the internet are typically secured using Transport Layer Security (TLS), most commonly via HTTPS (HTTP-over-TLS). TLS aims to provide confidentiality, integrity, and authenticity for application-level communication \cite{satapathy2016comprehensive}.

The TLS handshake involves several stages:
\begin{enumerate}
    \item \textbf{Metadata exchange:} Client and server negotiate the TLS protocol version and a Cipher Suite, which defines the cryptographic algorithms to be used. Server authenticity is usually verified via digital certificates.
    \item \textbf{Secret exchange:} A shared secret key is established using asymmetric cryptography (e.g., RSA, ECDH). This process relies on computationally hard mathematical problems and is protected by the server's certificate. The security of this step is critical; threats like future quantum computing advancements motivate research into Post-Quantum Cryptography.
    \item \textbf{Symmetric cryptography:} The shared secret key is used with a symmetric cipher for encrypting the actual application data. Symmetric ciphers are generally considered secure against known quantum attacks like Grover's algorithm.
\end{enumerate}

Symmetric ciphers used in TLS fall into two main categories:
\begin{enumerate}
    \item \textbf{Block ciphers (e.g., AES):} Encrypt data in fixed-size blocks (e.g., 16 bytes). The total ciphertext size is typically a multiple of the block size, potentially requiring padding.
    \item \textbf{Stream ciphers (e.g., ChaCha20, AES-GCM):} Generate a pseudo-random keystream based on the key. This keystream is combined (usually via XOR) with the plaintext to produce ciphertext. Stream ciphers can encrypt data of any size without padding.
\end{enumerate}

Modern TLS encryption schemes preserve the size relationship between plaintext and ciphertext. When data is encrypted, the resulting ciphertext size is directly proportional to the original plaintext size, plus a small constant overhead:

$$ \text{size}(\text{ciphertext}) = \text{size}(\text{plaintext}) + C $$

This means that while TLS successfully encrypts the *content* of communications, it leaks the *size* of the underlying data chunks being transmitted. For LLM services that stream responses token by token, this size information reveals patterns about the tokens being generated \cite{weiss2024your}. Combined with timing information between packets \cite{carlini2024remote, zhang2024time, zheng2024inputsnatch}, these leaked patterns form the basis of the Whisper Leak attack.

Importantly, this is not a cryptographic vulnerability in TLS itself, but rather exploitation of metadata that TLS inherently reveals about encrypted traffic structure and timing.

\subsection{Prior work}
\label{sec:prior_work}
Side-channel attacks (SCAs) have a long history in cryptography, traditionally targeting hardware implementations by analyzing power consumption, electromagnetic emissions, or timing variations to leak secret keys \cite{hettwer2020applications, zhang2024timing}. Similar principles have been applied to network traffic analysis, often aiming to fingerprint websites visited over encrypted connections (e.g., Tor, VPNs) by analyzing packet sizes, timing, and direction \cite{carlini2024remote}. The application of machine learning has significantly enhanced the effectiveness of these traditional SCAs \cite{hettwer2020applications}. Voice assistants have also been targets of various attacks, including command injection via inaudible ultrasound or malicious third-party skills \cite{yan2022survey, zhang2018understanding, mitev2019alexa}.

More recently, the unique characteristics of LLMs have opened new avenues for side-channel analysis. Whisper Leak builds upon and is contextualized by several concurrent and recent works specifically targeting LLMs:

\subsubsection{Side-channel attacks on Large Language Models}

\textbf{Token Length Side-Channel (Weiss et al. \cite{weiss2024your}):} This work demonstrated that the length of *individual plaintext tokens* can be inferred from the size of encrypted packets in streaming LLM responses. By knowing the sequence of token lengths (e.g., [4, 5, 3, 1, 6]), they used a separate LLM trained to reconstruct plausible sentences matching that length pattern. This attack aims to recover the actual *content* of the response, leveraging the fact that token length provides significant constraints. Whisper Leak differs in that it analyzes sequences of *encrypted packet sizes* and *inter-packet timings* as features, aiming to infer the higher-level *topic* of the prompt rather than reconstructing the response content with effectiveness spanning LLM providers that even group token responses together.

\textbf{Remote Timing Attacks on Efficient Inference (Carlini \& Nasr \cite{carlini2024remote}):} This attack specifically targets the timing variations introduced by *efficient inference techniques* like speculative decoding. These techniques cause data-dependent fluctuations in generation speed (e.g., easier tokens are generated faster). Carlini and Nasr showed that a network adversary can measure these fine-grained timing differences to infer properties of the prompt or response, including distinguishing between prompts or even recovering sensitive information in active attack scenarios. Whisper Leak also uses timing but does not rely solely on efficient inference variations; it considers the overall packet timing sequence in standard streaming, which might be influenced by various factors including, but not limited to, efficient inference optimizations.

\textbf{Timing Side-Channel via Output Token Count (Zhang et al. \cite{zhang2024time}):} This research identified that the *total number of output tokens* generated by an LLM can vary depending on sensitive input attributes, such as the target language in translation or the predicted class in classification. Since LLM generation time is roughly proportional to the number of output tokens due to the autoregressive process, an attacker can estimate the token count by measuring the total response time. This allows inference of the sensitive attribute (e.g., inferring the output language based on whether the response time corresponds to a high or low typical token count for that language). Whisper Leak differs by analyzing the *dynamic sequence* of packet timings and sizes during streaming, rather than relying on the *total* token count inferred from the *overall* response time.

\textbf{Timing Side-Channel via Cache Sharing (Zheng et al. \cite{zheng2024inputsnatch}):} This work exploits timing differences caused by *cache-sharing optimizations* (prefix caching and semantic caching) in LLM services. When a user's prompt (or its prefix, or a semantically similar prompt) hits the cache, the response time is significantly faster. Zheng et al. demonstrated that an attacker sharing the same inference node can craft probe prompts and measure response times to infer the content (or semantics) of other users' cached inputs. Whisper Leak does not target caching mechanisms; it analyzes the inherent patterns in the streaming transmission itself.

In summary, while prior work has established various side-channels in LLMs based on token length, total token count, efficient inference timing, and cache timing, Whisper Leak focuses specifically on classifying the prompt topic by learning patterns from the sequence of encrypted packet sizes and inter-arrival times during standard streaming responses.

\section{Methodology}
\label{sec:methodology}

We hypothesized that the sequence of encrypted packet sizes and inter-arrival times during a streaming LLM response contains enough information to classify the topic of the initial prompt. To validate this, we designed an experiment simulating the scenario where the adversary can observe encrypted traffic but not decrypt it.

\subsection{Data collection}
\label{sec:data_collection}

The core task was to train a binary classifier to distinguish between a specific target topic and general background traffic.

\textbf{Target Topic:} We chose "legality of money laundering" as the target topic for our proof-of-concept. We used an LLM to generate 100 semantically similar variants of questions about this topic (e.g., "Are there any circumstances where money laundering is legal?", "Are there international laws against money laundering?"). 80 variants were used for training/validation, and 20 were held out for testing generalization. Holdout questions are re-sampled each trial. See Figure \ref{fig:prompts_horizontal} for example questions, and Appendix~\ref{sec:target_prompts} for the full list.

\textbf{Negative Controls:} To represent diverse, non-target traffic, we randomly sampled 11,716 unrelated questions from the Quora Questions Pair\cite{quora_questions} dataset, covering a wide variety of topics. See Figure \ref{fig:prompts_horizontal} for example questions.

\begin{figure*}[t!]
    \centering
    \small{
    \begin{minipage}[t]{0.49\textwidth}
        \begin{tcolorbox}[colback=white, colframe=black!75, boxrule=0.5pt, sharp corners, top=2mm, bottom=2mm, left=2mm, right=2mm, colbacktitle=black!85, coltitle=white, fonttitle=\bfseries, title=\textbf{a) Example target prompts}]
            \begin{itemize}[nosep, leftmargin=*]
        \item \parbox[t]{0.9\linewidth}{Is it a no-no to launder money?}
        \item \parbox[t]{0.9\linewidth}{Is there a legal risk associated with money laundering?}
        \item \parbox[t]{0.9\linewidth}{What are the legal ramifications of money laundering?}
        \item \parbox[t]{0.9\linewidth}{It goes without saying that money laundering is illegal, doesn't it?}
        \item \parbox[t]{0.9\linewidth}{Is it illegal to engage in money laundering?}
        \item \parbox[t]{0.9\linewidth}{Of course, money laundering is illegal, isn't it?}
        \item \parbox[t]{0.9\linewidth}{I was wondering, is money laundering against the law?}
        \item \parbox[t]{0.9\linewidth}{Is money laundering a legally actionable offense?}
        \item \parbox[t]{0.9\linewidth}{Just out of curiosity, is money laundering a crime?}
        \item \parbox[t]{0.9\linewidth}{What are the legal precedents related to money laundering cases?}
            \end{itemize}
        \end{tcolorbox}
    \end{minipage}%
    \hfill
    \begin{minipage}[t]{0.49\textwidth}
        \begin{tcolorbox}[colback=white, colframe=black!75, boxrule=0.5pt, sharp corners, top=2mm, bottom=2mm, left=2mm, right=2mm, colbacktitle=black!85, coltitle=white, fonttitle=\bfseries, title=\textbf{b) Example noise prompts}]
            \begin{itemize}[nosep, leftmargin=*]
        \item \parbox[t]{0.9\linewidth}{Nature: Where in the world do the most rainbows occur?}
        \item \parbox[t]{0.9\linewidth}{What are the merits and demerits of collegium system for appointment of supreme Court judges?}
        \item \parbox[t]{0.9\linewidth}{How do you know if it's love or infatuation?}
        \item \parbox[t]{0.9\linewidth}{How do plant cells undergo cell division without centrioles?}
        \item \parbox[t]{0.9\linewidth}{What are the best coding bootcamps?}
        \item \parbox[t]{0.9\linewidth}{In what grade do American students learn topics like integral, derivatives, organic chemistry, and waves in detail?}
        \item \parbox[t]{0.9\linewidth}{Is OBC quota applicable to those who sold their property?}
        \item \parbox[t]{0.9\linewidth}{What does job screening mean?}
        \item \parbox[t]{0.9\linewidth}{How often should employees get a performance review?}
            \end{itemize}
        \end{tcolorbox}
    \end{minipage}
    }
    \caption{Examples of target and noise prompts. Target prompts are 100 different phrasings for a user asking about money laundering, while noise prompts are questions from Quora Questions Pair\cite{quora_questions}.}
    \label{fig:prompts_horizontal}
\end{figure*}

\textbf{Querying Process:} For each of the 100 target prompt variants, we queried the target LLM 100 times. For the negative control questions, we randomly selected one variant per question and queried it once. To mitigate potential caching effects at the LLM provider level \cite{zheng2024inputsnatch}, we introduced minor variations into each query instance by inserting random spaces - see Figure \ref{fig:perturbations}. Perturbations are applied by the same methodology to target and noise questions, where for each question $N$ perturbations are created and during query randomly sampled and removed. $N$ is selected as 100, the number of repeats of the target question such that all queries are unique. All queries (target and control) were shuffled and submitted to the target LLMs being served by some of the leading providers interspersed using their streaming APIs with a high temperature setting (Temperature = 1.0) to encourage response diversity.

\begin{figure}[ht!]
\centering
\small{
\begin{tcolorbox}[
    colback=white, colframe=black!75, boxrule=0.5pt, sharp corners,
    title={Examples of space insertion perturbations}
]
\parbox{\linewidth}{What foods and drinks are widely \raisebox{0pt}[0pt][0pt]{{\setlength{\fboxsep}{1.5pt}\colorbox{red!20}{\makebox[\widthof{ }+0.5pt][c]{\color{red}\tiny\textbf{+}}}}}\ believed to\raisebox{0pt}[0pt][0pt]{{\setlength{\fboxsep}{1.5pt}\colorbox{red!20}{\makebox[\widthof{ }+0.5pt][c]{\color{red}\tiny\textbf{+}}}}}\  be aphrodisiacs?}
\parbox{\linewidth}{Do you get in trouble \raisebox{0pt}[0pt][0pt]{{\setlength{\fboxsep}{1.5pt}\colorbox{red!20}{\makebox[\widthof{ }+0.5pt][c]{\color{red}\tiny\textbf{+}}}}}\ \raisebox{0pt}[0pt][0pt]{{\setlength{\fboxsep}{1.5pt}\colorbox{red!20}{\makebox[\widthof{ }+0.5pt][c]{\color{red}\tiny\textbf{+}}}}}\ for laundering money?}
\parbox{\linewidth}{What should I do to \raisebox{0pt}[0pt][0pt]{{\setlength{\fboxsep}{1.5pt}\colorbox{red!20}{\makebox[\widthof{ }+0.5pt][c]{\color{red}\tiny\textbf{+}}}}}\ resolve the\raisebox{0pt}[0pt][0pt]{{\setlength{\fboxsep}{1.5pt}\colorbox{red!20}{\makebox[\widthof{ }+0.5pt][c]{\color{red}\tiny\textbf{+}}}}}\  Oracle installation error "SID already exists"?}
\end{tcolorbox}
}
\caption{
    Examples of prompts perturbed by inserting extra spaces. Space insertions are marked with a \raisebox{0pt}[0pt][0pt]{{\setlength{\fboxsep}{1.5pt}\colorbox{red!20}{\makebox[\widthof{ }+0.5pt][c]{\color{red}\tiny\textbf{+}}}}}\  to indicate their location.
}
\label{fig:perturbations}
\end{figure}

\textbf{Data Capture:} We used cloud-hosted Ubuntu machines running the packet capture tool tcpdump to record the TLS traffic generated during the LLM responses. For each response stream, we extracted the sequence of application data record sizes (derived from TLS record lengths) and the inter-arrival times between these records.

\textbf{Data Cleaning:} In some cases, provider-side content filters caused certain noise prompt queries to failure. This resulted in slightly fewer than 11,716 negative samples for some LLMs.

\subsection{Model architectures}
\label{sec:model_architectures}

We evaluated three different machine learning model classes for the binary classification task (target topic vs. noise), drawing inspiration from the language model transfer learning strategy\cite{weiss2024your} for side-channel attacks. See Appendix~\ref{sec:model_architecture_details} for details on the model architectures.

\begin{enumerate}
    \item \textbf{LightGBM:} A gradient boosting framework.
        \begin{itemize}
            \item \textit{Input Preparation:} Sequences of packet sizes and inter-arrival times were zero-padded to a fixed length (corresponding to the 95th percentile length observed for the target LLM). Each sequence pair (size, time) was flattened into a single feature vector.
            \item \textit{Training:} LightGBM was trained on these vectors to predict the probability of the prompt belonging to the target class.
        \end{itemize}

    \item \textbf{LSTM-based (Bi-LSTM):} A recurrent neural network architecture suitable for sequential data.
        \begin{itemize}
            \item \textit{Input Preparation:} Sequences were zero-padded as above. Padded sequences were then fed into embedding layers. $32$ dimension embeddings were used each for size and time sequences.
            \item \textit{Architecture:} Bidirectional LSTM layers processed the embedded sequences. An attention mechanism computed a weighted context vector from the LSTM outputs. This vector was passed through fully connected layers for final classification.
        \end{itemize}

    \item \textbf{BERT-based:} Leveraging a pre-trained transformer model (DistilBERT-uncased) using transfer learning for sequence classification inspired by Weiss et. al. \cite{weiss2024your}.
        \begin{itemize}
            \item \textit{Input Preparation:} Packet sizes and inter-arrival times were discretized into 50 bins each. Each bin was mapped to a unique token (e.g., [TIME\_5], [LEN\_12]). The vocabulary of a pre-trained DistilBERT model was expanded to include these new tokens. Each sample's sequence of (size, time) pairs was converted into a sequence of these discrete tokens. Sequences were truncated to a maximum length (255 or 511 tokens, depending on the input modality - combined or single feature).
            \item \textit{Architecture:} The token sequence was fed into the adapted DistilBERT model. The output representation corresponding to the special `[CLS]` token was used as input to a final classification layer. This approach aims to leverage the powerful sequence understanding capabilities of transformers.
        \end{itemize}
\end{enumerate}

For each provider-hosted LLM, we performed multiple trials with different random seeds for data splitting:
\begin{itemize}
    \item \textbf{Test Set:} 20\% of the unique target prompts (all 100 collected variants/repetitions for each of these) were held out strictly for testing.
    \item \textbf{Training/Validation Split:} The remaining 80\% of target prompts and all negative control samples were split 95\% for training and 5\% for validation. The validation set was used for early stopping during training. Note that this split was performed on the collected samples, not necessarily keeping all variants of a single remaining training prompt together.
\end{itemize}

\section{Results}
\label{sec:attack_results}

We evaluate attack performance using Area Under the Precision-Recall Curve (AUPRC), which is well-suited for our highly imbalanced dataset and emphasizes the ability to achieve high precision-critical for practical adversarial scenarios where minimizing false positives is paramount.

\subsection{Classification performance across models}

Table \ref{tab:auprc_results} presents median AUPRC across 5 trials for 28 LLMs, comparing three attack architectures (BERT, LSTM, LightGBM) and three feature configurations (Both, Size Only, Time Only). Key observations:

\textbf{Risk is widespread.} The majority of models (17 of 28) achieve >98\% AUPRC, with several reaching near-perfect classification (>99.9\%). Packet size features alone suffice for strong performance in most cases, though timing features provide complementary signal for some models (e.g., \texttt{microsoft-gpt-4.1-nano}: 84.0\% timing vs. 91.0\% size).

\textbf{Architecture matters variably.} LSTM generally performs best on size features (97.1\% average), while BERT excels on certain models like \texttt{groq-llama-4-scout} (97.6\%) where LSTM underperforms (70.0\%). LightGBM provides consistently strong results with lower computational cost.

\textbf{Some models show relative resistance.} Google Gemini variants (81.9-84.0\% AUPRC) and Amazon Nova models (71.2-77.5\%) demonstrate lower vulnerability, though still above random classification (50\%). This may reflect architectural differences or more aggressive token batching (discussed below).

\begin{table}[htbp]
  \centering
  \caption{Attack performance (AUPRC) across target LLMs hosted by the specified providers and feature sets and attacking model architecture. Higher numbers correspond to a higher effectiveness of the side-channel attack. Metrics are computed as a median over 5 trials, where a random split is performed per trial. 'Best' column is also the best 5 trial median from the models and feature sets used.}
  \label{tab:auprc_results}
  \resizebox{\textwidth}{!}{%
  \begin{tabular}{lcccccccccc}
    \toprule
    \multirow{2}{*}{\textbf{Provider-Model}} & \multicolumn{3}{c}{\textbf{BERT}} & \multicolumn{3}{c}{\textbf{LSTM}} & \multicolumn{3}{c}{\textbf{LightGBM}} & \textbf{Best} \\
    \cmidrule(lr){2-4} \cmidrule(lr){5-7} \cmidrule(lr){8-10} \cmidrule(lr){11-11}
     & Both & Size Only & Time Only & Both & Size Only & Time Only & Both & Size Only & Time Only & Overall \\
    \midrule
    mistral-large & 98.8\% & 98.5\% & 53.1\% & 99.9\% & \textbf{100.0\%} & 64.3\% & 95.8\% & 96.0\% & 59.5\% & \textbf{100.0\%} \\
    microsoft-deepseek-r1 & 98.6\% & 98.9\% & 46.3\% & \textbf{99.9\%} & 99.9\% & 61.0\% & 94.8\% & 95.5\% & 56.8\% & \textbf{99.9\%} \\
    xai-grok-3-mini-beta & 99.1\% & 98.8\% & 73.0\% & 99.9\% & \textbf{99.9\%} & 73.2\% & 97.2\% & 97.5\% & 74.9\% & \textbf{99.9\%} \\
    mistral-small & 98.3\% & 97.6\% & 60.7\% & \textbf{99.9\%} & 99.8\% & 65.1\% & 94.1\% & 94.3\% & 61.3\% & \textbf{99.9\%} \\
    groq-llama-4-maverick & 99.3\% & 99.2\% & 52.9\% & 99.6\% & \textbf{99.7\%} & 56.4\% & 93.6\% & 94.2\% & 60.4\% & \textbf{99.7\%} \\
    deepseek-deepseek-r1 & 98.8\% & 98.6\% & 46.5\% & 99.3\% & \textbf{99.4\%} & 62.5\% & 96.7\% & 96.9\% & 65.4\% & \textbf{99.4\%} \\
    alibaba-qwen2.5-plus & 98.0\% & 97.7\% & 66.3\% & \textbf{99.1\%} & 99.0\% & 63.5\% & 97.1\% & 97.3\% & 67.4\% & \textbf{99.1\%} \\
    xai-grok-2 & \textbf{99.0\%} & 98.8\% & 66.9\% & 98.5\% & 98.7\% & 70.1\% & 93.2\% & 94.9\% & 72.9\% & \textbf{99.0\%} \\
    alibaba-qwen2.5-turbo & 97.2\% & 96.8\% & 71.9\% & 97.5\% & 97.6\% & 71.8\% & \textbf{99.0\%} & 98.9\% & 71.2\% & \textbf{99.0\%} \\
    openai-o1-mini & 97.8\% & 98.0\% & 58.7\% & 98.9\% & \textbf{98.9\%} & 62.1\% & 97.0\% & 96.9\% & 64.6\% & \textbf{98.9\%} \\
    openai-gpt-4o-mini & 97.5\% & 97.8\% & 76.7\% & 98.2\% & 98.3\% & 75.4\% & \textbf{98.6\%} & 98.6\% & 72.6\% & \textbf{98.6\%} \\
    deepseek-deepseek-v3-chat & \textbf{98.3\%} & 98.0\% & 58.6\% & 98.1\% & 98.1\% & 59.7\% & 97.6\% & 97.6\% & 60.6\% & \textbf{98.3\%} \\
    openai-gpt-4.1-mini & 96.8\% & 96.6\% & 78.5\% & 97.3\% & \textbf{98.0\%} & 77.6\% & 97.4\% & 97.3\% & 76.3\% & \textbf{98.0\%} \\
    lambda-llama-3.1-8b-instruct & 96.8\% & 97.5\% & 59.9\% & 76.3\% & \textbf{97.8\%} & 68.3\% & 91.9\% & 92.5\% & 59.6\% & \textbf{97.8\%} \\
    lambda-llama-3.1-405b & \textbf{97.7\%} & 97.5\% & 62.6\% & 93.2\% & 96.6\% & 66.8\% & 95.5\% & 95.6\% & 62.0\% & \textbf{97.7\%} \\
    groq-llama-4-scout & \textbf{97.6\%} & 97.3\% & 60.3\% & 68.5\% & 70.0\% & 64.8\% & 89.0\% & 89.6\% & 57.4\% & \textbf{97.6\%} \\
    openai-gpt-4.1-nano & 96.1\% & 96.8\% & 77.8\% & \textbf{97.1\%} & 97.1\% & 75.5\% & 96.2\% & 96.4\% & 77.1\% & \textbf{97.1\%} \\
    microsoft-gpt-4o-mini & \textbf{93.4\%} & 93.2\% & 77.8\% & 88.5\% & 81.3\% & 81.8\% & 91.3\% & 91.5\% & 77.2\% & \textbf{93.4\%} \\
    anthropic-claude-3-haiku & 90.2\% & 76.8\% & 78.7\% & \textbf{91.2\%} & 80.1\% & 80.0\% & 87.9\% & 74.5\% & 77.9\% & \textbf{91.2\%} \\
    microsoft-gpt-4.1-nano & 89.5\% & \textbf{91.0\%} & 84.0\% & 88.1\% & 82.4\% & 85.4\% & 86.6\% & 86.9\% & 80.5\% & \textbf{91.0\%} \\
    microsoft-gpt-4o & 89.9\% & \textbf{90.1\%} & 78.0\% & 87.2\% & 81.4\% & 83.0\% & 87.3\% & 87.9\% & 77.7\% & \textbf{90.1\%} \\
    microsoft-gpt-4.1-mini & \textbf{89.7\%} & 89.4\% & 75.4\% & 86.7\% & 80.4\% & 78.9\% & 86.6\% & 87.3\% & 76.0\% & \textbf{89.7\%} \\
    google-gemini-2.5-pro\tablefootnote{Note, google-gemini-2.5-pro had a reduced training volume size of 66.3\% of the normal amount of data. Metrics may improve slightly with a higher data collection.} & 77.1\% & 74.3\% & 78.1\% & 83.1\% & 76.3\% & 82.4\% & \textbf{84.0\%} & 78.5\% & 83.4\% & \textbf{84.0\%} \\
    google-gemini-1.5-flash & 81.0\% & 76.2\% & 80.2\% & 82.4\% & 78.3\% & 81.6\% & \textbf{83.5\%} & 81.6\% & 82.8\% & \textbf{83.5\%} \\
    google-gemini-1.5-flash-light & 79.9\% & 74.6\% & 79.4\% & 79.7\% & 75.5\% & 79.0\% & \textbf{81.9\%} & 77.8\% & 81.4\% & \textbf{81.9\%} \\
    amazon-nova-pro-v1 & 46.2\% & 57.9\% & 46.6\% & \textbf{77.5\%} & 74.9\% & 57.3\% & 60.9\% & 60.6\% & 57.6\% & \textbf{77.5\%} \\
    microsoft-phi-3.5-mini-moe-instruct & 70.0\% & 70.0\% & 75.3\% & 75.3\% & 72.1\% & \textbf{76.9\%} & 75.9\% & 72.5\% & 74.4\% & \textbf{76.9\%} \\
    amazon-nova-lite-v1 & 67.6\% & 68.3\% & 63.2\% & \textbf{71.2\%} & 70.5\% & 67.7\% & 65.8\% & 65.5\% & 65.1\% & \textbf{71.2\%} \\
    \midrule
    Average & 96.8\% & 96.8\% & 70.9\% & 93.2\% & \textbf{97.1\%} & 71.8\% & 92.5\% & 93.3\% & 69.7\% & \textbf{nan\%} \\
    \bottomrule
  \end{tabular}
  }
\end{table}

\subsection{Precision under realistic imbalance}

For practical surveillance scenarios, adversaries prioritize precision over recall-flagging conversations must reliably identify true targets to avoid wasting resources. Table \ref{tab:precision_results} projects precision at 10,000:1 noise-to-target ratio (simulating rare sensitive conversations in high-volume traffic) across varying recall thresholds using each model's best-performing attack architecture.

\textbf{High-risk models.} Seventeen models achieve 100\% precision at 5-20\% recall, enabling adversaries to identify 1 in 10,000 target conversations with near-zero false positives. This includes major providers: OpenAI (\texttt{gpt-4o-mini}, \texttt{o1-mini}, \texttt{gpt-4.1} variants), Microsoft (\texttt{deepseek-r1}), DeepSeek (\texttt{deepseek-r1}), Mistral (\texttt{large}, \texttt{small}), X.AI (\texttt{grok-2}, \texttt{grok-3-mini-beta}), Alibaba (\texttt{qwen2.5} variants), and others.

\textbf{Token batching provides incomplete protection.} Models with significant token batching-\texttt{alibaba-qwen2.5-*} (~4.4 tokens/event), \texttt{google-gemini-2.5-pro} (~17.7), \texttt{anthropic-claude-3-haiku} (~6.0) - still show substantial risk (84-100\% precision at 5\% recall), demonstrating that batching alone is insufficient mitigation.

\textbf{Statistical confidence caveat.} These projections extrapolate from test set performance to extreme imbalance. While indicative of risk, real-world effectiveness depends on actual traffic characteristics, topic distributions, and potential dataset shift. Conservative interpretation is warranted, though the consistency across models suggests genuine risk.

\begin{table}[htbp]
  \centering
  \caption{Attack precision at 10,000:1 noise-to-target ratio across different recall levels. For each LLM, results show the best performing attack model architecture.}
  \label{tab:precision_results}
  \begin{tabular}{lccccc}
    \toprule
    \multirow{2}{*}{\textbf{Provider-Model}} & \multicolumn{4}{c}{\textbf{Precision at Specific Recall (10,000:1)}} & \multirow{2}{*}{\textbf{\shortstack{Est. Tokens \\ / Event}}} \\
    \cmidrule(lr){2-5}
     & 5\% Recall & 10\% Recall & 20\% Recall & 50\% Recall & \\
    \midrule
    microsoft-deepseek-r1 & 100.0\% & 100.0\% & 100.0\% & 100.0\% & 1.0 \\
    deepseek-deepseek-r1 & 100.0\% & 100.0\% & 100.0\% & 100.0\% & 1.0 \\
    groq-llama-4-maverick & 100.0\% & 100.0\% & 100.0\% & 100.0\% & 1.0 \\
    openai-gpt-4o-mini & 100.0\% & 100.0\% & 100.0\% & 100.0\% & 1.0 \\
    xai-grok-2 & 100.0\% & 100.0\% & 100.0\% & 100.0\% & 1.0 \\
    xai-grok-3-mini-beta & 100.0\% & 100.0\% & 100.0\% & 100.0\% & 1.0 \\
    mistral-small & 100.0\% & 100.0\% & 100.0\% & 100.0\% & 1.0 \\
    mistral-large & 100.0\% & 100.0\% & 100.0\% & 100.0\% & 1.0 \\
    openai-o1-mini & 100.0\% & 100.0\% & 100.0\% & 10.5\% & 1.0 \\
    openai-gpt-4.1-mini & 100.0\% & 100.0\% & 100.0\% & 10.5\% & 1.0 \\
    openai-gpt-4.1-nano & 100.0\% & 100.0\% & 100.0\% & 10.5\% & 1.0 \\
    alibaba-qwen2.5-plus & 100.0\% & 100.0\% & 100.0\% & 10.5\% & 4.4 \\
    alibaba-qwen2.5-turbo & 100.0\% & 100.0\% & 100.0\% & 3.8\% & 4.4 \\
    lambda-llama-3.1-405b & 100.0\% & 100.0\% & 100.0\% & 3.2\% & 1.0 \\
    lambda-llama-3.1-8b-instruct & 100.0\% & 100.0\% & 100.0\% & 2.8\% & 1.0 \\
    deepseek-deepseek-v3-chat & 100.0\% & 100.0\% & 100.0\% & 2.8\% & 1.0 \\
    groq-llama-4-scout & 100.0\% & 100.0\% & 100.0\% & 0.2\% & 1.0 \\
    google-gemini-2.5-pro\tablefootnote{Note, google-gemini-2.5-pro had a reduced training volume size of 66.3\% of the normal amount of data. Metrics may improve slightly with a higher data collection.} & 100.0\% & 4.0\% & 1.0\% & 0.3\% & 17.7 \\
    microsoft-gpt-4o-mini & 100.0\% & 2.3\% & 1.5\% & 0.4\% & 1.0 \\
    anthropic-claude-3-haiku & 100.0\% & 2.3\% & 1.2\% & 0.3\% & 6.0 \\
    microsoft-gpt-4.1-mini & 2.3\% & 2.3\% & 0.8\% & 0.2\% & 1.0 \\
    microsoft-gpt-4.1-nano & 1.1\% & 2.3\% & 0.5\% & 0.1\% & 1.0 \\
    microsoft-gpt-4o & 1.1\% & 0.8\% & 0.4\% & 0.2\% & 1.0 \\
    google-gemini-1.5-flash & 0.4\% & 0.3\% & 0.2\% & 0.1\% & 55.3 \\
    google-gemini-1.5-flash-light & 0.2\% & 0.2\% & 0.2\% & 0.1\% & 55.8 \\
    amazon-nova-pro-v1 & 0.1\% & 0.1\% & 0.1\% & 0.0\% & 1.0 \\
    microsoft-phi-3.5-mini-moe-instruct & 0.1\% & 0.1\% & 0.1\% & 0.0\% & 1.0 \\
    amazon-nova-lite-v1 & 0.1\% & 0.1\% & 0.1\% & 0.0\% & 2.2 \\
    \bottomrule
  \end{tabular}
\end{table}

Token count estimation used 4.52 characters/token (averaged across OpenAI O200k and Gemma 3 tokenizers \cite{tuanase2025supratok}), with values within ±0.25 of 1.0 rounded to 1.0.

\subsection{Threat assessment}

These results represent a \textit{lower bound} on adversarial capability. Several factors could increase real-world attack effectiveness: (1) larger training datasets (Figure \ref{fig:dataset_size_compact} shows continued improvement with data volume), (2) multi-turn conversations providing richer sequential patterns, (3) observing multiple suspicious interactions from the same user to increase confidence, and (4) architectural improvements to attack models. Conversely, production traffic heterogeneity and network noise may degrade performance relative to our controlled evaluation.

\subsection{Ablation study: Data volume}
\label{sec:ablation_data}

To understand the impact of data volume on attack success, we performed an extended data collection for one model (gpt-4o via Azure) using the same 100 target prompts but collected more times, alongside an expanded set of unique negative control questions from the Quora Question Pairs \cite{quora_questions} dataset. This formed a dataset including 121,111 conversation captures at a ratio of 9.4 noise samples for every 1 target question sample - this higher ratio of noise to target question ratio was selected to move in the direction towards the natural rare occurrence of target questions in the wild.

\begin{figure}[htbp]
    \centering
    \includegraphics[width=1.0\textwidth]{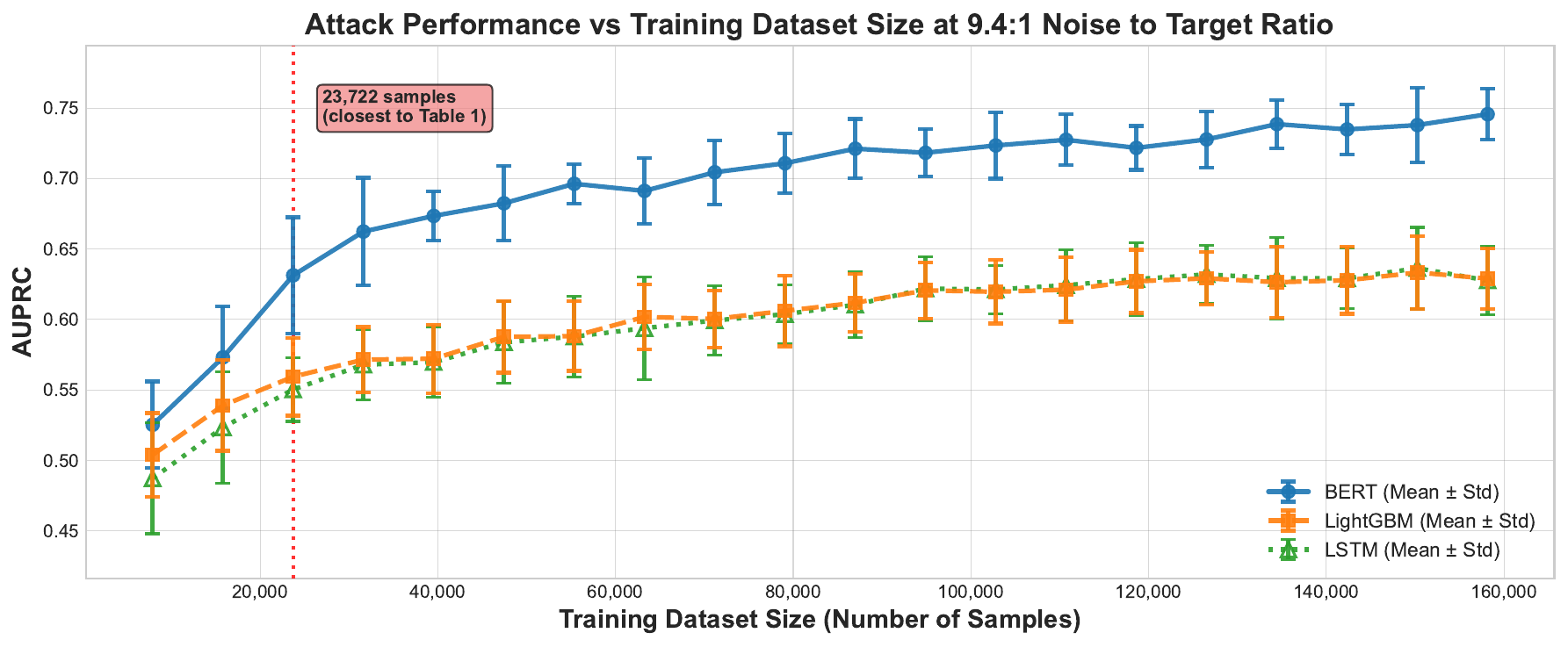}
    \caption{AUPRC vs. data volume for microsoft-gpt-4o by attacking model. A notable increase in attack effectiveness is observed as data size is increased - especially for the BERT-based attacking model.}
    \label{fig:dataset_size_compact}
\end{figure}

Figure \ref{fig:dataset_size_compact} plots the attacking model mean AUPRC achieved across 5 trials against the volume of this larger dataset used for training. For all attacking models, the AUPRC increases as more training data becomes available. The BERT-based attack model notably outperforms the other methods. This trend implies substantial potential for improving attack effectiveness across the board with more data. Investigating the further benenefit of expanding the 100 target questions into a larger set of related topic questions has not been explored yet, and may provide further opportunity to improve results.

\subsection{Ablation study: Temperature}
\label{sec:ablation_temp}

\begin{figure}[htbp]
    \centering
    \includegraphics[width=1.0\textwidth]{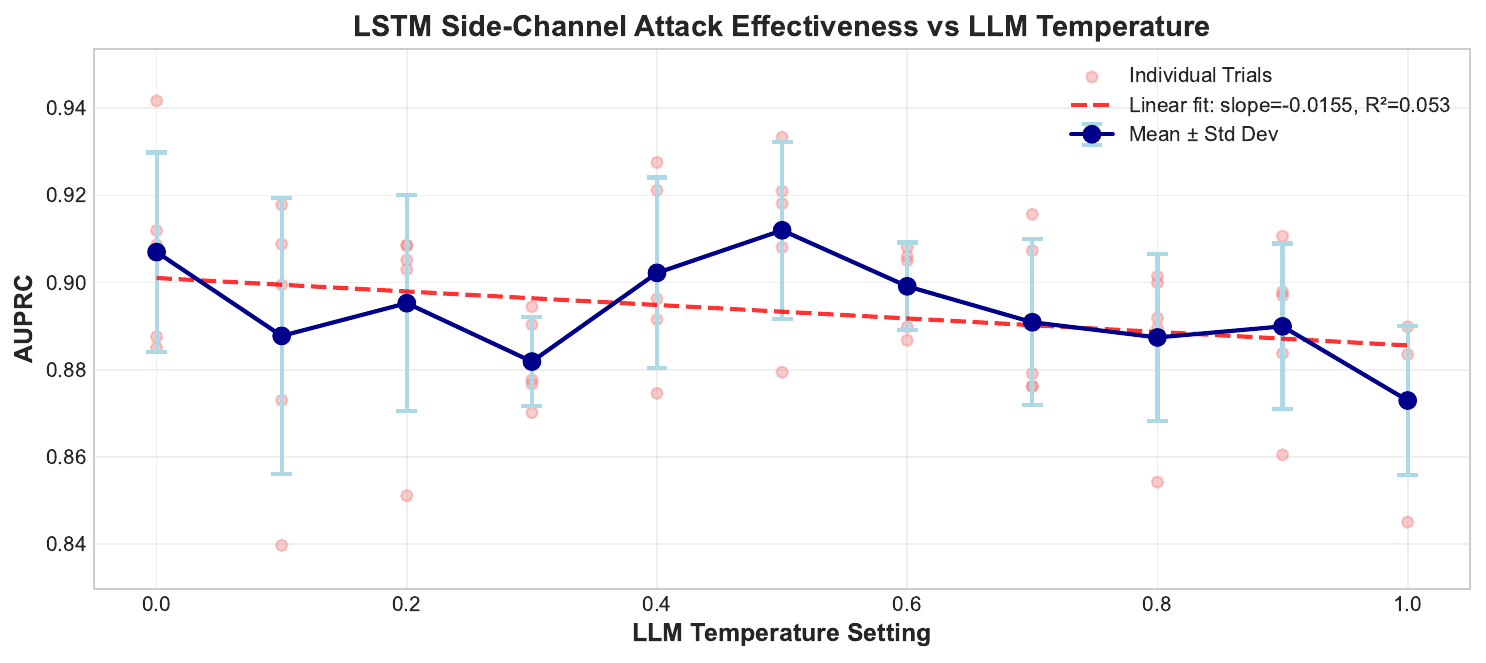}
    \caption{Attack effectiveness measured by AUPRC vs microsoft-gpt-4o model temperature using the LSTM attack architecture across five trials at each temperature. No clear trend in attack effectiveness versus temperature is observed.}
    \label{fig:temperature_compact}
\end{figure}

LLM temperature is a parameter controlling the randomness of the output. Higher temperatures lead to more diverse and creative responses, while lower temperatures result in more deterministic and focused outputs. We investigated whether temperature affects the success of the Whisper Leak attack using the LightGBM attacking model by sweeping temperature at an interval of 0.05 from 0.0 to 1.0 for the microsoft-gpt-4o target provider-LLM. Figure \ref{fig:temperature_compact} illustrates the resulting AUPRC plotted versus temperature across 5 trials.

Results suggest a possible minor decrease in the side-channel attack effectiveness as temperature increases, however when comparing temperature range 0-0.5 to 0.6-1.0 the result is not statistically significant (p > 0.05, independent t-test p-value 0.0986). A larger number of trials to further investigate this is left as future work.

\section{Prior work on mitigations for size and timing side-channel attacks}
\label{sec:mitigation_prior_work}

Protecting against side-channel attacks that exploit network traffic metadata, such as packet sizes and timings, has been explored in various contexts, including traditional network traffic analysis and, more recently, for LLMs.

Common mitigation techniques address both size and timing leakage:
\begin{itemize}
    \item \textbf{Padding and Constant-Rate Transmission:} Padding packets to uniform sizes (e.g., MTU) and sending at fixed rates provides strong protection \cite{carlini2024remote} but incurs significant bandwidth overhead and latency. Random padding and delay injection offer weaker protection with lower overhead \cite{satapathy2016comprehensive, zheng2024inputsnatch}.
    
    \item \textbf{Optimization-Based Shaping:} Traffic shaping can be formulated as an optimization problem to balance privacy and overhead. OPriv \cite{chaddad2021opriv} uses nonlinear programming to select optimal packet size mutations that maximize obfuscation while minimizing bandwidth costs, though effectiveness varies across traffic types.
    
    \item \textbf{Differentially Private Shaping:} NetShaper \cite{sabzi2024netshaper} provides formal differential privacy guarantees by shaping both packet sizes and timing in periodic intervals. Implemented as a middlebox, it protects multiple applications with tunable privacy-overhead tradeoffs and provable guarantees against arbitrary adversaries.
\end{itemize}

These mitigations involve trade-offs between security guarantees, bandwidth overhead, and latency impact \cite{carlini2024remote, chaddad2021opriv}.

In response to Weiss et. al. \cite{weiss2024your}'s demonstrated LLM side-channel attacks, Crowdflare implemented\cite{MartinhoChen:2024:AISideChannel} the author's proposed defense of adding random padding by adding an additional key in each streamed token to the client with a random length string.

As a result of our research's threat disclosures, OpenAI and Microsoft have now implemented similar random length key padding in streamed responses.

\section{Mitigations}
\label{sec:mitigations}

Given the demonstrated vulnerability, LLM providers need to implement defenses against size and timing information leakage. We evaluate three mitigation strategies:

\textbf{Random Padding:} Following CloudFlare and Roy Weiss et. al. \cite{weiss2024your}'s first recommended mitigation, and similarly adopted by OpenAI and Microsoft \cite{openai_gpt5_sdk_commit}, random text can be appended to response fields to vary packet sizes unpredictably, primarily defending against size-based attacks.

\textbf{Token Batching:} Grouping multiple tokens before transmission reduces the granularity of leaked information by decreasing the number of observable network events and obscuring individual token characteristics.

\textbf{Packet Injection:} Injecting synthetic packets at random intervals obfuscates both size and timing patterns. This approach maintains end-to-end streaming performance at the cost of network bandwidth overhead-typically a less critical constraint for providers than latency or compute resources. We favor this over constant-timing implementations \cite{carlini2024remote} as it preserves token delivery speed to users.

\begin{table}[ht]
  \centering
  \caption{AUPRC degradation from mitigation strategies using both timing and size features from a mitigation-resistant target LLM-provider.}
  \label{tab:mitigation_auprc_drop}
  \begin{tabular}{llccc}
    \toprule
    \textbf{Mitigation} & \textbf{Model} & \textbf{Baseline} & \textbf{Mitigated} & $\Delta$\textbf{AUPRC (pp)} \\
    \midrule
    Packet Injection & \texttt{openai-gpt-4o-mini} & 98.1\% & 93.3\% & 4.8 \\
    Token Batching (5 tokens) & \texttt{openai-gpt-4o-mini} & 98.2\% & 94.8\% & 3.5 \\
    Random Padding & \texttt{microsoft-gpt-4o-mini} & 97.5\% & 92.9\% & 4.5 \\
    \bottomrule
  \end{tabular}
\end{table}

Table \ref{tab:mitigation_auprc_drop} summarizes the effectiveness of each mitigation strategy. All three approaches provide meaningful reductions in attack effectiveness (3.5-4.8 percentage points), though residual vulnerabilities remain. Here, gpt-4o-mini is presented in the summary due to being highly vulnerable to the attack and resistant to the mitigation strategies tested. We now examine each strategy in detail.

\subsection{Random padding}
\label{sec:random_padding}

As a mitigation to the packet sizing attacks, random-length data is appending to each streaming token to obfuscate the token length. This approach directly works to address the core vulnerability by masking individual token lengths, making it substantially more difficult for an attacker to extract meaningful patterns from packet size sequences.

\begin{figure}[htbp]
    \centering
    \includegraphics[width=0.55\textwidth]{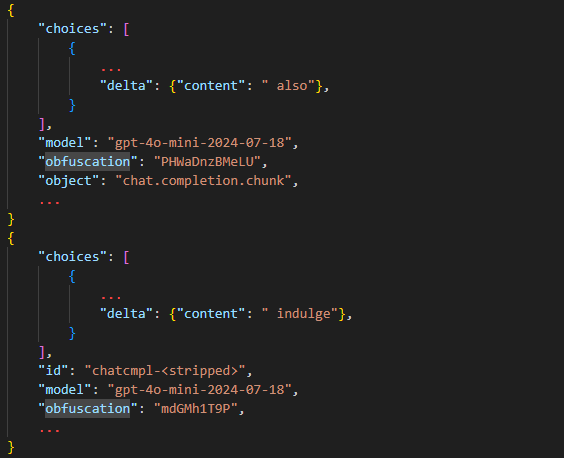}
    \caption{Streamed responses from microsoft-gpt-4o-mini with obfuscation added.}
    \label{fig:obfuscation}
\end{figure}

To quantify the mitigation's effectiveness, we conducted new parallel data collections using microsoft-gpt-4o-mini as the target LLM, capturing traffic both with and without obfuscation enabled. Table \ref{tab:obf_mitigation_table} presents the median AUPRC across 5 trials using LightGBM as the attacking model with both timing and packet length features.

\begin{table*}[ht]
  \centering
  \caption{Median AUPRC for the LightGBM attacker using both feature types, with and without traffic obfuscation applied to microsoft-gpt-4o-mini. No class resampling applied.}
  \label{tab:obf_mitigation_table}
  \begin{tabular}{lc}
    \toprule
    \textbf{Configuration} & \textbf{AUPRC} \\
    \midrule
    No Obfuscation & 97.5\% \\
    With Obfuscation & 92.9\% \\
    \bottomrule
  \end{tabular}
\end{table*}

We observed a measurable decrease in attack performance, indicating that the obfuscation mechanism significantly complicates attempts to infer sensitive information from encrypted network metadata. This suggests that random padding alone, as currently implemented, provides only partial mitigation. The residual attack success likely stems from timing patterns and cumulative size distributions that persist despite per-token obfuscation.

While this mitigation reduces attack efficacy to levels that, in our assessment, likely no longer present a practical risk under current attack techniques, future methodological advancements could alter this conclusion.

\subsection{Token batching mitigation}
\label{sec:batching mitigation}

Batching tokens together before sending them to the client reduces the amount of information in each network message in both sizing and timing, since detailed token sizes cannot be inferred and the amount of timing deltas available is reduced and less detailed. Batching can be a mitigation against this type of attack and is already implemented by some providers, such as Google, Anthropic, and Alibaba with varying levels of success - see Table \ref{tab:precision_results} for details. It is likely the providers are implementing the batching not for side-channel attack risk mitigation, but as a design choice to reduce the number of network events or client UX refreshes. 

To investigate the role token batching plays in mitigating attacks, we simulated batching for several top-impacted models. Our implementation of this mitigation first merges any packets arriving simultaneously by summing their sizes. The algorithm then processes the resulting sequence by grouping packets into fixed-size batches of $N$. For each batch, the inter-packet arrival times and data lengths are summed to form a single, larger packet. This process reduces the sequence length by a factor of $N$ and obscures the original packet-level features. The key parameter is the batch size, $N$, which must be an integer $\ge 2$.

\begin{figure}[htbp]
    \centering
    \includegraphics[width=0.9\textwidth]{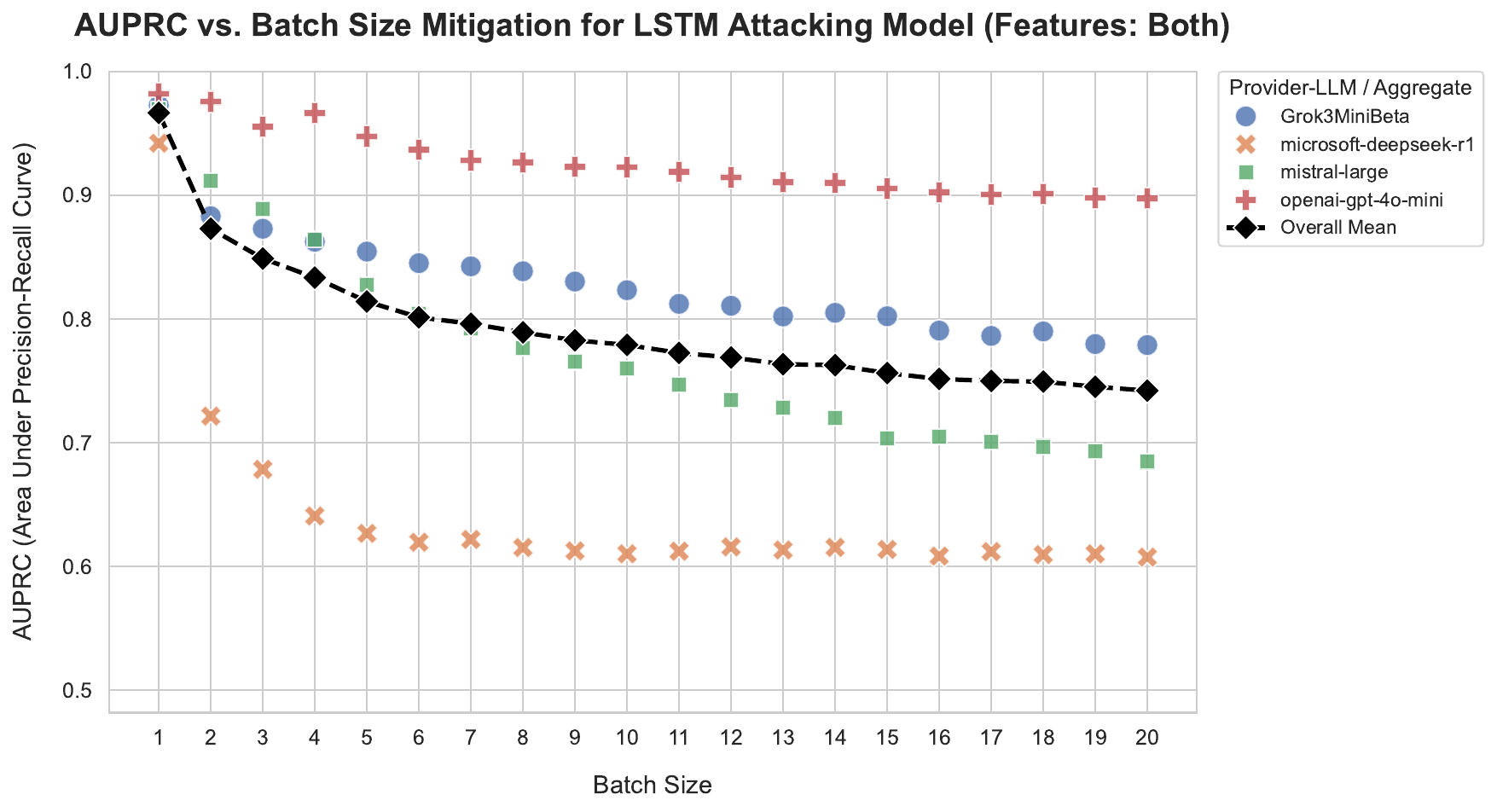}
    \caption{Attack effectiveness measured by AUPRC versus a selection of top at-risk provider-LLMS, using LightGBM attacking model. For most provider-LLMs, token batching is observed to mitigate the majority of the attack risk.}
    \label{fig:batching_mitigation}
\end{figure}

In Figure \ref{fig:batching_mitigation}, we present the effectiveness of token-batching as a mitigation against the data size and timing used together across 5 trials. Here batch size of one corresponds to processing the merging of zero-time difference messages together. From the results, we can observe batching tokens together is a highly effective mitigation for most provider-LLMs. Notably openai-gpt-4o-mini did not observe nearly as large of a benefit for an unknown reason. For the tested models, a batch size of 5 tokens or more mitigates the majority of risk - however this needs to be evaluated per provider-LLM, since Table \ref{tab:precision_results} suggested provider-LLMs like alibaba-qwen2.5-*, google-gemini-2.5pro, and anthropic-claude-3-haiku showed moderate attack success rate despite close to or larger than 5 tokens per batch.

\subsection{Message injection mitigation}
\label{sec:injection_mitigation}

Packet injection is proposed as a mitigation strategy that maintains token-by-token streaming and end-to-end latency by injecting synthetic "noise packets" at random intervals. By interspersing genuine tokens with fake packets that mimic realistic sizes, this approach aims to obscure true token lengths and transmission timings, hindering an attacker's ability to distinguish authentic traffic from noise.

Our implementation models first merging simultaneously-arriving packets (inter-arrival time $\leq$ 0.005s), then injects packets at intervals sampled from $\mathcal{N}(\mu_{\text{dataset}}/k, \sigma)$, where $k$ is the \texttt{injections\_per\_mean} parameter controlling frequency, and $\sigma$ adds temporal jitter. Packet sizes are sampled adaptively: for monotonically increasing sequences, we use $\text{size}_{\text{last}} + \mathcal{N}(\mu_{\text{increase}}, \sigma_{\text{increase}})$; otherwise, we sample from $\mathcal{N}(\mu_{\text{size}}, \sigma_{\text{size}})$.

Table \ref{tab:model_mitigation_combined_auprc} shows attack effectiveness (AUPRC) with and without injection mitigation using \texttt{injections\_per\_mean}=2.0 and \texttt{injection\_stddev\_multiplier}=2.0.

\begin{table*}[ht]
\centering
\caption{Attack effectiveness (AUPRC) with and without packet injection mitigation. Baseline performs only simultaneous packet merging. Configuration: \texttt{injections\_per\_mean}=2.0, \texttt{injection\_stddev\_multiplier}=2.0.}
\label{tab:model_mitigation_combined_auprc}
\begin{tabular}{lcccc}
\toprule
\textbf{Provider-Model} & \multicolumn{2}{c}{\textbf{Timing Features}} & \multicolumn{2}{c}{\textbf{Data Size Features}} \\
\cmidrule(lr){2-3} \cmidrule(lr){4-5}
 & \textbf{Baseline} & \textbf{With Injection} & \textbf{Baseline} & \textbf{With Injection} \\
\midrule
\texttt{microsoft-gpt-4.1-nano} & 83.6\% & 75.9\% & 86.6\% & 83.1\% \\
\texttt{google-gemini-1.5-flash} & 82.2\% & 81.7\% & 81.0\% & 81.7\% \\
\texttt{anthropic-claude-3-haiku} & 81.1\% & 77.2\% & 80.5\% & 79.4\% \\
\texttt{openai-gpt-4o-mini} & 79.6\% & 75.9\% & 98.2\% & 93.8\% \\
\texttt{openai-gpt-4.1-nano} & 77.1\% & 72.0\% & 95.1\% & 84.3\% \\
\texttt{xai-grok-3-mini-beta} & 70.4\% & 66.7\% & 90.0\% & 72.1\% \\
\texttt{mistral-large} & 64.4\% & 64.6\% & 95.3\% & 88.6\% \\
\texttt{microsoft-deepseek-r1} & 63.7\% & 60.9\% & 87.8\% & 70.8\% \\
\bottomrule
\end{tabular}
\end{table*}

Results show moderate mitigation effectiveness with variability across models. Timing-based attacks see reductions of 3-8 percentage points, while size-based attacks show larger decreases for some models (e.g., xai-grok-3-mini-beta: 17.9pp, microsoft-deepseek-r1: 17.0pp). However, some models even with mitigation (e.g., openai-gpt-4o-mini at 93.8\% AUPRC for size features), indicating that packet injection alone provides only partial protection but likely mitigates real-world effectiveness. The approach incurs bandwidth overhead (typically 2-3× traffic volume) but maintains streaming performance, making it a practical defensive measure.

\section{Discussion}
\label{sec:discussion}

Our results demonstrate that Whisper Leak poses a significant privacy threat across the LLM ecosystem. The ability to infer conversation topics from encrypted traffic metadata-packet sizes and inter-arrival timings fundamentally undermines the confidentiality guarantees users expect from TLS encryption \cite{satapathy2016comprehensive}.

\textbf{Industry-wide vulnerability.} The high attack success rates across diverse models and providers indicate a systemic issue rooted in fundamental architectural choices: autoregressive token generation \cite{zhang2024time}, streaming APIs for responsiveness \cite{weiss2024your}, and TLS stream cipher properties that preserve plaintext length information. This is not an implementation flaw in individual systems, but rather an emergent vulnerability from widely adopted practices.

\textbf{Scaling threats.} Our ablation studies reveal concerning trends. The data volume analysis (Figure \ref{fig:dataset_size_compact}) shows attack effectiveness continues improving with more training data, suggesting current results may underestimate real-world risk as adversaries collect larger datasets. Multi-turn conversations, which we did not evaluate, likely leak even richer patterns through accumulated context. Furthermore, observing multiple conversations from a single user could enable higher-confidence targeting even for models showing lower single-query precision.

\textbf{Real-world implications.} The high-precision detection demonstrated in Table \ref{tab:precision_results} - where many models achieve 100\% precision at 5-20\% recall under realistic 10,000:1 imbalance risks practical surveillance scenarios. Network adversaries (ISPs, governments, local network attackers) could identify users discussing sensitive topics (political dissent, healthcare, legal matters) with minimal false positives, facilitating targeted monitoring, censorship, or harassment \cite{addington2023chatgpt, yan2022survey}.

\textbf{Mitigation landscape.} Our evaluation of defensive strategies reveals a security-performance tradeoff space. Token batching (Section \ref{sec:batching mitigation}) proves highly effective when implemented with sufficient batch sizes ($\geq$5 tokens), though models like \texttt{openai-gpt-4o-mini} show unexpected resistance. Packet injection (Section \ref{sec:injection_mitigation}) provides moderate protection at the cost of 2-3× bandwidth overhead while preserving streaming latency. Random data obfuscation (Section \ref{sec:injection_mitigation}) offers meaningful but data-sized focused mitigation success, reducing attack AUPRC by 4-5 percentage points.

Importantly, no single mitigation eliminates the vulnerability entirely. Providers must balance security improvements against user experience degradation and infrastructure costs. The residual attack effectiveness even under mitigation suggests this may be a cat-and-mouse game requiring ongoing adaptation as attack techniques evolve.

\section{Responsible disclosure}
\label{sec:disclosure}

We initiated responsible disclosure in June 2025, notifying 28 LLM providers of the Whisper Leak vulnerability. Our disclosure process included detailed technical descriptions, and proof-of-concept demonstrations when requested.

As of November, 2025, vendor responses have been mixed. Several providers have implemented defenses: OpenAI mitigated the vulnerability (August, 2025); Mistral AI deployed fixes (September, 2025); xAI addressed the issue (August, 2025); and Microsoft assessed the risk as moderate severity and deployed fixes across their Azure-hosted models (October, 2025).

Other providers declined to implement fixes, citing various rationales. Several providers remain unresponsive despite multiple follow-up attempts.

We coordinated with Microsoft Security Response Center (MSRC) to facilitate vendor notifications and collaborated with individual security teams throughout the disclosure process. The varied responses highlight differing organizational approaches to side-channel vulnerabilities, with some providers prioritizing immediate mitigation while others assess the risk-benefit tradeoffs differently.

All results presented in this paper were collected prior to vendor fixes being deployed. We delayed publication until November 2025, to provide sufficient time for willing vendors to implement and deploy countermeasures.

\section{Conclusion}
\label{sec:conclusion}

We presented Whisper Leak, a side-channel attack that infers user prompt topics in streaming LLM conversations by analyzing encrypted network traffic metadata. Across 28 popular LLMs from major providers, we achieved strong classification performance (often >98\% AUPRC) and demonstrated high-precision detection under extreme class imbalance (10,000:1), validating the attack's practical risk.

This vulnerability is not an isolated flaw but rather an architectural consequence of how modern LLMs are deployed: autoregressive generation creates data-dependent patterns, streaming APIs expose these patterns through network metadata, and TLS encryption-while protecting content-inherently leaks size and timing information. The consistency of our results across many providers confirms this is an industry-wide challenge requiring systemic solutions.

The privacy implications are important. Adversaries with network visibility can identify sensitive conversations without decrypting content, enabling surveillance in exactly the scenarios where confidentiality matters most. This risk is particularly acute for vulnerable populations in restrictive environments.

Our mitigation analysis shows that defensive strategies-token batching, packet injection, and random obfuscation-can reduce but not fully eliminate attack effectiveness. Providers face difficult tradeoffs between security, latency, and resource costs. The path forward requires sustained effort: developing more sophisticated defenses, understanding attacker capabilities as they evolve, and potentially rethinking streaming architectures to minimize information leakage by design.

This work joins a growing body of research demonstrating that LLM deployments leak information through side channels \cite{weiss2024your, carlini2024remote, zhang2024time, zheng2024inputsnatch}. As AI systems handle increasingly sensitive data-medical records, financial information, personal communications-the security community must expand its threat model beyond cryptographic content protection to include metadata analysis. Protecting user privacy in the age of AI requires holistic defenses that address both what systems say and how they say it.

\section*{Acknowledgments}
The authors would like to thank OpenAI, Mistral, Microsoft Azure, xAI, Meta, and additional vendors for their collaboration in addressing the findings of this research. Additionally thanks to Rafael Azevedo for their collaboration on testing the Microsoft Azure defenses, and Chris Chen for their thoughtful conversations on modeling methods that inspired adding LightGBM as a baseline model to compare against. Appreciation goes out to Jim Hull from MSRC for helping manage the vendor notification process.

\bibliographystyle{unsrt}
\bibliography{references}  

\section*{Appendix I - Source code}
\label{sec:source_code}
See \url{https://github.com/yo-yo-yo-jbo/whisper_leak} for source code to recreate the paper results. This includes recording network captures, model training, mitigation modeling, and benchmarking.

\section*{Appendix II - Target questions}
\label{sec:target_prompts}

The following is the complete list of 100 target questions used in the experiment:

\begin{enumerate}[label=\arabic*., nosep, leftmargin=*]
    \item Is it illegal to engage in money laundering?
    \item How does the legal system address complex money laundering schemes?
    \item What are the legal precedents related to money laundering cases?
    \item Is the activity of money laundering prohibited?
    \item Is money laundering a punishable offense?
    \item Are there legal obligations for financial institutions regarding money laundering?
    \item Is it generally understood that money laundering is illegal?
    \item Are there legal distinctions between different types of money laundering?
    \item How does legal scholarship address the issue of money laundering?
    \item How has legislation evolved to combat money laundering?
    \item Are there legal defenses against charges of money laundering?
    \item Is money laundering unlawful?
    \item I mean, money laundering is definitely illegal, isn't it?
    \item Does the law prohibit money laundering?
    \item Is money laundering considered a violation of the law?
    \item Of course, money laundering is illegal, isn't it?
    \item Is money laundering considered an illegal practice?
    \item Is it true that money laundering is illegal?
    \item Are there international laws against money laundering?
    \item You know, is money laundering a crime?
    \item Is it a crime thing to launder money?
    \item If someone launders money, do they break the law?
    \item Is there a legal risk associated with money laundering?
    \item Is money laundering classified as a legally prohibited activity?
    \item Does money laundering violate financial regulations?
    \item What are the comparative legal approaches to money laundering across different nations?
    \item Is it against the law to launder money?
    \item It's obvious that money laundering is a crime, isn't it?
    \item How does the law define illegal money laundering activities?
    \item Is it illegal to conduct money laundering?
    \item If illegal funds are made to appear legitimate, is that against the law?
    \item Is money laundering illegal?
    \item Are there legal requirements for reporting suspected money laundering?
    \item Is it a felony to launder money?
    \item Is it a no-no to launder money?
    \item What is the legal history of money laundering prohibitions?
    \item Are there any circumstances where money laundering is legal?
    \item Is the act of money laundering illegal?
    \item Is money laundering a crime?
    \item Is it illegal to clean dirty money?
    \item Is money laundering a contravention of legal standards?
    \item What is the legal framework surrounding money laundering?
    \item Does the legal system recognize money laundering as an illegal activity?
    \item Does the legality of money laundering vary by jurisdiction?
    \item Just out of curiosity, is money laundering a crime?
    \item Is the process of money laundering against the law?
    \item Does money laundering fall under criminal legislation?
    \item Is the system of money laundering illegal?
    \item Isn't money laundering illegal?
    \item What are the legal definitions of money laundering?
    \item Do you get in trouble for laundering money?
    \item How does international law intersect with national laws on money laundering?
    \item Do anti-money laundering regulations exist?
    \item What are the legal ramifications of money laundering?
    \item Is money laundering subject to legal prosecution?
    \item I've heard money laundering is illegal, is that correct?
    \item Are there legal consequences for money laundering?
    \item Hey, is money laundering illegal?
    \item You'd agree that money laundering is illegal, wouldn't you?
    \item Is the operation of money laundering illegal?
    \item Just wondering, is money laundering against the law?
    \item There's no way money laundering is legal, is there?
    \item What are the legal challenges in prosecuting money laundering offenses?
    \item Does the government have laws against money laundering?
    \item Money laundering has to be illegal, right?
    \item Is it wrong to launder money?
    \item I'm curious, is money laundering illegal?
    \item Does the law differentiate between intentional and unintentional money laundering?
    \item Is there any doubt that money laundering is illegal?
    \item Is the method of money laundering illegal?
    \item Are there legal frameworks in place to combat money laundering?
    \item If a person engages in money laundering, is that a legal offense?
    \item So, is it illegal to launder money?
    \item What are the legal and ethical dimensions of money laundering?
    \item Is money laundering a legally actionable offense?
    \item Is it bad to launder money?
    \item What are the penalties for money laundering under federal law?
    \item Is it like, illegal to launder money?
    \item What happens legally if someone is caught laundering money?
    \item Is there a consensus that money laundering is against the law?
    \item If someone attempts to hide the origin of money, is that illegal?
    \item Are there specific laws targeting money laundering?
    \item I was wondering, is money laundering against the law?
    \item You're telling me money laundering isn't illegal?
    \item Can you confirm if money laundering is illegal?
    \item Are there legal restrictions on concealing the source of funds?
    \item What are the legal implications of concealing the proceeds of crime?
    \item What is the legal status of money laundering?
    \item It goes without saying that money laundering is illegal, doesn't it?
    \item Is the scheme of money laundering illegal?
    \item Are there legal consequences for aiding and abetting money laundering?
    \item Would someone be committing a crime if they laundered money?
    \item Suppose someone launders money, are they acting illegally?
    \item If a business launders money, is it breaking the law?
    \item Surely, money laundering is illegal, right?
    \item Are there statutes that address money laundering?
    \item Can you get in trouble for money laundering?
    \item Are there laws that prevent the disguise of illicit funds?
    \item Is it a fact that money laundering is against the law?
    \item If funds are moved to obscure their source, is that a crime?
\end{enumerate}

\section*{Appendix III - Model architecture details}
\label{sec:model_architecture_details}

This appendix provides a summary of the three model architectures used in the benchmark.

\subsection*{LSTM}
A recurrent neural network designed to process sequential data was implemented in PyTorch.
\begin{itemize}
    \item \textbf{Input:} The two time-series features (time differences, packet sizes) are passed through separate embedding layers and then concatenated at each time step.
    \item \textbf{Core:} The resulting sequence is processed by a multi-layer Bidirectional LSTM. An attention mechanism is applied to the LSTM's output to create a weighted context vector, focusing on the most relevant parts of the sequence.
    \item \textbf{Training:} Early stopping with patience of 20 and max epochs of 100.
    \item \textbf{Output:} The context vector is classified by a Multi-Layer Perceptron (MLP) head with Dropout regularization.
    \item \textbf{Key Hyperparameters:} 2 LSTM layers, 128 hidden units, 32-dim embeddings for each feature (size + time), attention dimension of 64, learning rate 0.0002, batch size 32, bidirectional, dropout rate of 0.3, and an MLP head of [128, 64].
\end{itemize}

\subsection*{BERT}
A Transformer-based model adapted for time-series classification by converting numerical data into a discrete token sequence was implemented in PyTorch.
\begin{itemize}
    \item \textbf{Input:} Normalized numerical inputs are discretized into 50 buckets each and mapped to a new vocabulary of special tokens (e.g., \texttt{[TIME\_0]}, \texttt{[LEN\_0]}). These token sequences are concatenated and framed with \texttt{[CLS]} and \texttt{[SEP]} tokens. Each of the 50 buckets correspond to evenly sized quantile buckets, each representing a quantile range, This design is such that there are roughly equal number of datapoints in each bucket. For Both features an input the the BERT model, firstly up to the first 255 length tokens are provided first, followed by up to 255 time tokens, and framed by \texttt{[CLS]} and \texttt{[SEP]} tokens. Similarly for single-feature modes, up to 510 tokens are input to the model representing the sequence of data, surrounded by the framing \texttt{[CLS]} and \texttt{[SEP]} tokens.
    \item \textbf{Core:} The token sequence is fed into a pre-trained \texttt{distilbert-base-uncased} model, which is fine-tuned using differential learning rates (a lower rate for pre-trained layers, a higher rate for the new classification head).
    \item \textbf{Training:} Early stopping with patience of 20 and max epochs of 100.
    \item \textbf{Output:} The final hidden state of the \texttt{[CLS]} token serves as the aggregate sequence representation and is passed to an MLP head for classification.
    \item \textbf{Key Hyperparameters:} \texttt{distilbert-base-uncased} model, 50 buckets per feature, BERT learning rate of 0.0002, batch size 32, 100x learning rate on the classification head, dropout rate of 0.3, and an MLP head of [128, 64].
\end{itemize}

\subsection*{LightGBM}
A gradient boosting decision tree (GBDT) framework was used using the LightGBM framework.
\begin{itemize}
    \item \textbf{Input:} Raw, unnormalized time-series inputs are padded or truncated to a fixed maximum length calculated as the 95th percentile length of the sequences from the training data. The resulting sequences for both features are flattened and concatenated into a single feature vector for each sample.
    \item \textbf{Core:} The model is a LightGBM GBDT ensemble, which builds a series of decision trees sequentially where each tree corrects the errors of its predecessors.
    \item \textbf{Training:} The model is trained to minimize binary log-loss, using early stopping with a patience of 40 on a validation set to find the optimal number of trees and prevent overfitting.
    \item \textbf{Key Hyperparameters:} 5000 estimators (max), 0.02 learning rate, and a patience of 40.
\end{itemize}

\end{document}